\documentclass[aps,prl,reprint,groupedaddress]{revtex4-1}
\usepackage[pdftex]{graphicx}
\usepackage{amsmath}

\begin{document}

\title{Sub-10 fs pulses tunable from 480 to 980 nm from a NOPA pumped by a Yb:KGW source}

\author{Matz Liebel}
\author{Christoph Schnedermann}
\author{Philipp Kukura}
\email{Corresponding author: philipp.kukura@chem.ox.ac.uk}
\affiliation{Physical and Theoretical Chemistry Laboratory, Department of Chemistry, University of Oxford, South Parks Road, Oxford OX1 3QZ, United Kingdom}

\begin{abstract}
We describe two noncollinear optical parametric amplifier (NOPA) systems pumped by either the second (515 nm) or the third (343 nm) harmonic of an Yb:KGW amplifier, respectively. Pulse durations as short as 6.8 fs are readily obtained by compression with commercially available chirped mirrors. The availability of both second and third harmonic for NOPA pumping allows for gap-free tuning from 520 to 980 nm. The use of an intermediate NOPA to generate seed light at 780 nm extends the tuning range of the third-harmonic pumped NOPA towards 450 nm. 
\end{abstract}

\maketitle


Noncollinear optical parametric amplifiers (NOPAs) \cite{Wilhelm1997,Cerullo1998} have become the standard tool for the generation of tunable femtosecond pulses from the UV \cite{Homann2012} to the IR \cite{Brida2008}. While NOPA designs for Ti:Sapphire (800 nm) laser systems are readily available \cite{Cerullo2003} delivering pulses with sub-5 fs pulse duration \cite{Baltuska2002}, only one scheme for Ytterbium (1030 nm) based laser systems has been reported \cite{Homann2008}. The work highlighted the enormous potential of Ytterbium lasers by demonstrating continuous tunability over more than one octave (440-1000 nm) with pulse durations as short as 20 fs. Here, we report the layout of two NOPAs for generation of sub-10 fs pulses in the 480-650 nm (VIS) as well as in the 680-980 nm (NIR) spectral regions, with continuous tunability ranging from 450-980 nm. The NIR-NOPA is pumped by the second, the VIS-NOPA by the third harmonic of an Yb:KGW (1030 nm) based laser system respectively.

\begin{figure}[htbp]
\centerline{\includegraphics[width=1\columnwidth]{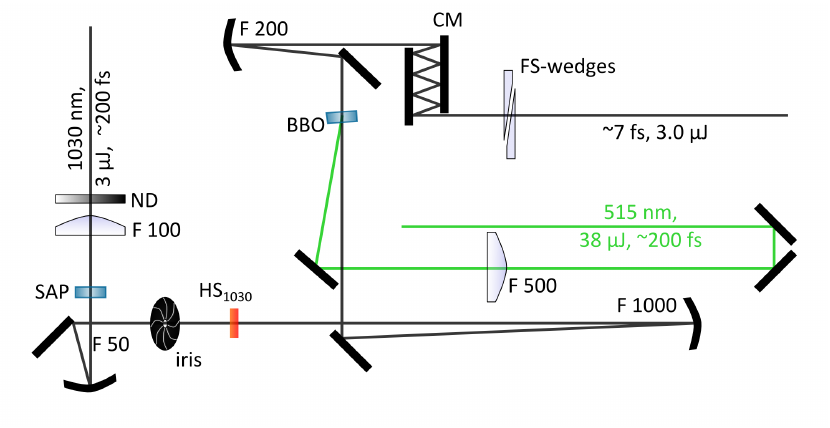}}
\caption{\label{fig:fig1}Schematic of the second harmonic (515 nm) pumped NIR-NOPA: ND, variable neutral density filter; SAP, 3 mm sapphire plate (Foctek Photonics); $\mathrm{HS}_{1030}$, 1 mm fused silica, 0$^{\circ}$ AOI, 1030 nm harmonic separator (EKSMA Optics); BBO, $\beta$-barium borate 2.0 mm $\theta$ = 23.5$^{\circ}$ type 1 (Foctek Photonics); CM, 680-980 nm chirped mirror compression GDD and TOD optimised (Layertec); FS-wedges, 2$^{\circ}$ apex angle fused silica prims pair (Layertec).}
\end{figure}

The design of the NIR-NOPA pumped by the second harmonic (515 nm) largely follows that of a standard Ti:Sapphire NOPA (FIG. \ref{fig:fig1}). We focused 3.0 $\mu$J of the fundamental with a 100 mm focal length lens into a 3 mm sapphire plate to generate the white light (WL) seed. Given the excellent mode quality of the fundamental, irising prior WL generation, did not improve the mode or stability of the WL. To keep the seed chirp to a minimum, we collimated the emerging WL with a 50 mm focal length mirror before selecting the spectrally uniform central 20 \% (1 mm diameter) by means of an iris. We found that a hard apertured WL seed markedly improved both the mode as well as the compressibility of the final output, while only marginally reducing the output power. A harmonic separator (EKSMA Optics) removed residual fundamental which otherwise undergoes strong amplification. The group velocity dispersion added by the sapphire plate and the harmonic separator resulted in a WL duration that matched that of the 515 nm pump, thus leading to improved conversion efficiency during the amplification process. 38 $\mu$J of the pump and the WL seed were overlapped within the $\beta$-barium borate (BBO, $\theta$ = 23.5$^{\circ}$) crystal at an internal angle of 2.5$^{\circ}$ in order to ensure broadband phase matching. The pump was focused 8.5 cm, the seed 5.0 cm before the BBO with the crystal to lens distances being 59.5 cm and 49.0 cm respectively, resulting in a  peak intensity of approximately 150 $\mathrm{GW}/\mathrm{cm}^{2}$ at the crystal. We found that reducing the peak intensity at the expense of output power below the commonly employed 200 $\mathrm{GW}/\mathrm{cm}^{2}$ mark lead to improved compressibility and mode quality. Moving the collimation mirror after WL generation allowed us to adjust the seed spot size to be 2.5 times smaller than the pump spot in the crystal and lead to spatially uniform amplification. 

\begin{figure}[htbp]
\centerline{\includegraphics[width=1\columnwidth]{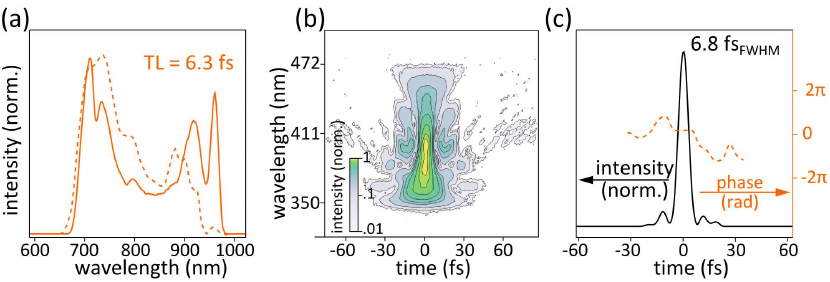}}
\caption{\label{fig:fig2}Performance of the NIR-NOPA: (a) Spectra of the amplified 3.0 $\mu$J signal obtained with 515 nm pumping (solid) and SHG-FROG retrived spectrum (dashed). (b) Corresponding SHG-FROG. (c) Retrieved temporal profile and phase.}
\end{figure}

The output power after single stage amplification amounted to 3.0 $\mu$J corresponding to a conversion efficiency of 8$\%$ covering 130 THz over the 680-980 nm spectral range (FIG. \ref{fig:fig2}a). Pulse compression is  achieved using dispersion compensating chirped mirrors (Layertec) in combination with a pair of fused silica wedge-prisms. We note that optimal pulse compression is only possible with group-delay and third-order-dispersion compensating mirrors \cite{Zavelani-Rossi2001}. The pulse duration retrieved by second harmonic generation frequency resolved optical gating (SHG-FROG) \cite{Kane1993} in a 10 $\mu$m BBO was 6.8 fs (FIG. \ref{fig:fig2}b-c), very close to the transform limit of 6.3 fs. The residual oscillatory phase is most likely caused by small sinusoidal modulations of the chirped mirrors phase profile.

\begin{figure}[htbp]
\centerline{\includegraphics[width=1\columnwidth]{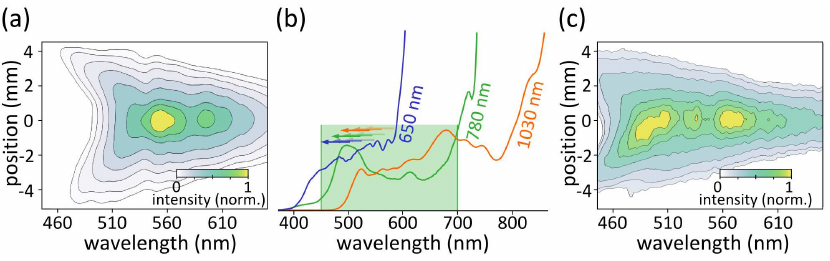}}
\caption{\label{fig:fig4}780 nm seed pulse and WL properties: (a) Spectral distribution as a function of spatial position within the WL obtained with a 1030 nm seed. (b) Comparison of WL spectra obtained for different seed wavelengths: 1030 nm (orange), 780 nm (green) and 650 nm (blue). (c) Spectral distribution as a function of spatial position within the WL obtained with a 780 nm seed.}
\end{figure}

\begin{figure}[htbp]
\centerline{\includegraphics[width=1\columnwidth]{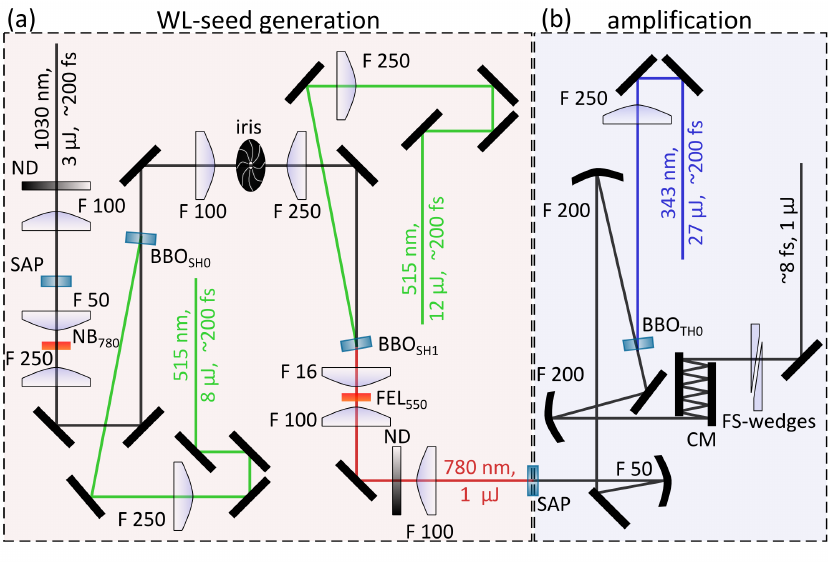}}
\caption{\label{fig:fig3}Schematic of the VIS-NOPA: ND, variable neutral density filter; SAP, 3 mm sapphire plate (Foctek Photonics); $\mathrm{NB}_{780}$, 780 nm dielectric 10 nm FWHM bandpass filter (FB780-10, Thorlabs); $\mathrm{BBO}_{\mathrm{SH0}}$, 2.0 mm $\theta$ = 26.5$^{\circ}$ type 1 (Foctek Photonics); $\mathrm{BBO}_{\mathrm{SH1}}$, 2.0 mm $\theta$ = 23.5$^{\circ}$ type 1 (Foctek Photonics);  $\mathrm{BBO}_{\mathrm{TH0}}$, 2.0 mm $\theta$ = 37$^{\circ}$ type 1 (Foctek Photonics);    $\mathrm{FEL}_{550}$, 550 nm dielectric longpass filter (FEL0550, Thorlabs); $\mathrm{CM}$, 480-650 nm chirped mirror compression GDD and TOD optimized (Layertec); FS-wedges, 2$^{\circ}$ apex angle fused silica prism pair (Layertec).}
\end{figure}

In order to obtained femtosecond pulses in the visible, amplification with the 343 nm pump is necessary, allowing for broadband phase matching of the 450-700 nm spectral region. Continuous tunability in this range has been demonstrated previously \cite{Homann2008}, yielding sub-20 fs pulses over the full visible region. When amplifying a bandwidth supporting sub-12 fs pulses, however, we found it exceedingly difficult to achieve levels of compression close to the transform limit, especially when moving towards the bluer part of the spectrum. We attribute this to the spectrally non-uniform spatial distribution for wavelengths below 510 nm within the WL (FIG. \ref{fig:fig4}a) leading to severe space-time coupling within the amplified beam. Blue shifting the WL seed and thus the overall WL spectrum (FIG. \ref{fig:fig4}b) as well as the onset of the non uniformity within the WL itself circumvents this problem (FIG. \ref{fig:fig4}c). The generation of a high quality mode WL at 780 nm allows us to fully exploit the broad amplification bandwidth available with 343 nm pumping, while simultaneously maintaining excellent compressibility. By shifting the WL seed even further (650 nm) it is possible to obtain uniform amplification down to 400 nm.
FIG. \ref{fig:fig3} shows a schematic of the VIS-NOPA with an intermediate stage to generate the blue-shifted WL seed. Such schemes have previously been applied to cover the tuning gap in Ti:Sapphire NOPAs \cite{Manzoni2012}.  The setup is separated into two independently operating NOPA systems which can be adjusted individually. The WL seed generation stage (FIG. \ref{fig:fig3}a) outputs a pulse centered at 780 nm with a bandwidth of ~200 $\mathrm{cm}^{-1}$ that is subsequently used for WL generation in a 3 mm sapphire plate. In the amplification stage (FIG. \ref{fig:fig3}b), the WL obtained with the 780 nm seed was amplified to yield tunable pulses in the visible spectral region, by means of 343 nm pumping. 

While higher band gap materials such as fused silica or $\mathrm{CaF}_{2}$ \cite{Bradler2009} can be used for the generation of a blue shifted WL, material translation is necessary in order to avoid bulk damage. At 10 kHz operation we found it difficult to generate a stable and spectrally constant $\mathrm{CaF}_{2}$ WL, and almost impossible to maintain it over the course of several hours. The blue shifted  WL seed in combination with photostable sapphire is the far more stable and easier to maintain alternative.

Within the WL generation stage we selected the desired output wavelength (780 nm) from the 1030 nm pumped sapphire WL using a narrowband filter (FIG. \ref{fig:fig3}a). The spectral selection and narrowing of the seed light not only matches its pulse duration to the one of the pump but further eliminates any risk of spectral detuning during NOPA alignment.  It is thus possible to optimize the two-stage NOPA using only the output power as feedback. After spectral filtering, the seed and the pump pulses were overlapped at a crossing angle of 0.5$^{\circ}$ within the first BBO crystal. BBO to lens distances of 27.5 cm for the pump and 26.5 cm for the seed resulted in respective focus positions of 2.5 cm and 1.5 cm before the crystal face. After amplification, we obtained 15 nJ pulses centered at 780 nm. We hard aperture selected the center of the amplified pulse, which ultimately lead to a Gaussian mode profile after 2-stage amplification. The recollimated seed and pump pulses were then focused into the second BBO, crossing at an internal angle of 0.5$^{\circ}$. We set the pump focus position 3.0 cm and the seed 2.0 cm before the crystal with the BBO to lens distances being 28 cm and 25.5 cm respectively. 

The collimated output obtained after 2-stage amplification is 1 $\mu$J centered at 780 nm with an elliptical Gaussian (2:1) mode. We decided not to correct for the ellipticity, as could be done by means of cylindrical telescope, as no improvement in the generated WL was observed. When comparing the WL obtained with the 780 nm to the one obtained at 1030 nm (FIG. \ref{fig:fig4}b,c), the desired blue shift is apparent, moving the spectrum into the fully phase-matched amplification region. Even though identical output power could be obtained with a single stage NOPA we opted for the slightly more complex 2-stage option as we found that the improved mode quality drastically benefited the WL quality and stability. The  WL obtained in this way exhibits a spatially uniform spectral distribution leading to readily compressible pulses after amplification.

\begin{figure}[htbp]
\centerline{\includegraphics[width=1\columnwidth]{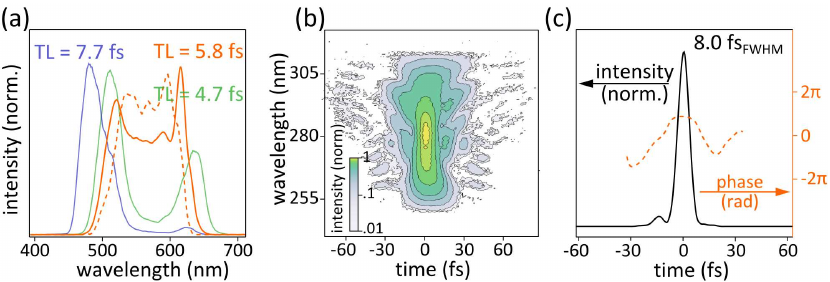}}
\caption{\label{fig:fig5}Performance of the VIS-NOPA pumped at 343 nm: (a) Amplified output pulses obtained with 343 nm pumping. Pulse used for compression (orange) and pulses highlighting the tuning capability with a 2 mm BBO (green, blue) as well as the SHG-FROG retrieved pulse spectrum (dashed, orange). (b) SHG-FROG obtained for the orange pulse shown in (a). (c) Retrieved temporal profile and phase.}
\end{figure}

We consecutively amplified the 780 nm WL by means of the third harmonic (343 nm) employing a pump to seed internal angle of 4.2$^{\circ}$. Focusing optics to crystal distances were set to 29 cm and 28 cm for pump and probe, with the focal points being 4 cm and 3 cm before the BBO, respectively. A short focal length mirror (F=50 mm) allows for fine adjustment of the probe focusing conditions within the BBO. In this configuration only the center of the WL is amplified, avoiding the spectrally non uniform regions at the rim of the WL. We opted for a peak intensity of only 150 $\mathrm{GW}/\mathrm{cm}^{2}$ resulting in excellent long term stability at the cost of lower conversion efficiency. Spectra representative of recollimated output pulses with around 1 $\mu$J intensity are shown in FIG. \ref{fig:fig5}a. 

A combination of chirped mirrors (Layertec) and a fused silica wedge prism pair were used for compression of the output pulse. A pulse duration of 8.0 fs was retrieved from the SHG-FROG trace recorded in 10 $\mu$m BBO (FIG. \ref{fig:fig5}b-c). Our tuning bandwidth for compressible pulses is currently limited by the mirror’s chirp compensation range (480-650 nm). Bluer pulses, as the ones shown in FIG. \ref{fig:fig5}a, were successfully compressed to sub-9 fs using a combination of grating and a deformable mirror. It is further possible to achieve spectrally flat broadband amplification in the 450-700 nm spectral region by exchanging the 2 mm for a 1 mm BBO, hence increasing the phase matching bandwidth \cite{Cerullo2003}.

Both the VIS- and NIR-NOPAs presented in this work exhibit limited tuning capabilities as essentially the full phase matching bandwidth is amplified. This is mainly due to the relatively long pulse durations (200 fs) of the pump pulses. We found that the easiest way of reducing the pulse bandwidth, while maintaining the output power, is to introduce additional material into the WL seed before amplification. Insertion of 2-3 mm of YAG chirp the WL to a duration that allows for amplification of spectrally Gaussian pulses tunable over the full respective phase matching ranges. The transform limit of pulses generated in this fashion is on the order of 10 fs and compression is achieved by increasing the number of bounces on the chirped mirrors. The tuning range can be further extended into the UV (350-450 nm) and MUV (250-300 nm) by frequency doubling the NOPA output in a thin BBO crystal or by means of more advanced schemes such as achromatic frequency doubling \cite{Baum2004}.

In conclusion, we presented two NOPA schemes pumped with the second and third harmonics of an Ytterbium based amplifier system outputting sub-10 fs pulses in the near infrared and visible regions, respectively. Both NOPA designs were optimized to match the needs of ultrafast spectroscopists, keeping both complexity and maintenance requirements at a minimum while maximizing long term stability. We highlighted the potential of Yb-based amplification by readily exceeding amplification bandwidths only accessible with Ti:Sapphire based NOPA schemes. The continuous tunability demonstrated here allows us to achieve full amplification of 450-980 nm pulses by coupling VIS- and NIR-NOPA\cite{Manzoni2012}.

We thank Maximilian Bradler, Dr. Daniele Brida and Dr. Alexander Weigel for support during the development of the NOPAs.

\bibliography{references}
\end{document}